\documentclass[aps,prl,reprint,groupedaddress,amsmath]{revtex4-1}

\usepackage{graphicx}
\begin{document}

%\preprint{}

\title{Nonlinear optomechanics in the stationary regime}

% Please use programthe appropriate macro foreach each type of information

\author{C. Doolin}
\author{B.D. Hauer}
\author{P.H. Kim}
\author{A.J.R. MacDonald}
\author{H. Ramp}
\author{J.P. Davis}
\email[]{jdavis@ualberta.ca}
%\thanks{}
\affiliation{Department of Physics, University of Alberta, T6G 2E1 Edmonton, AB, Canada}


\date{\today}

\begin{abstract}

We have observed nonlinear transduction of the thermomechanical motion of a nanomechanical resonator when detected as laser transmission through a sideband unresolved optomechanical cavity.   Nonlinear detection mechanisms are of considerable interest as special cases allow for quantum nondemolition measurements of the mechanical resonator's energy.   We investigate the origin of the nonlinearity in the optomechanical detection apparatus and derive a theoretical framework for the nonlinear signal transduction, and the optical spring effect, from both nonlinearities in the optical transfer function and second order optomechanical coupling.  By measuring the dependence of the linear and nonlinear signal transduction -- as well as the mechanical frequency shift -- on laser detuning from optical resonance, we provide estimates of the contributions from the linear and quadratic optomechanical couplings.
\end{abstract}

\maketitle

%%%%%%%%%%%%%%%%%%%
%
%\section{Introduction \label{sec.intro}}
%
%%%%%%%%%%%%%%%%%%%

Cavity optomechanics has resulted in new levels of extremely precise displacement transduction \cite{ Teufel09, Anetsberger10} of ultrahigh frequency resonators \cite{Eichenfield09}.  This has created much interest in pursuing quantum measurements \cite{Schwab05} of nanomechancial devices \cite{OConnell2010,Safavi-Naeini12,Palomaki13}, as well as dynamical back action cooling \cite{Schliesser08,Groblacher09,Chan11,Teufel11}.  

One of the most fundamental, and as of yet unattained, quantum measurements that could be performed is that of the quantized energy eigenstates of a nanomechanical resonator (as has been demonstrated with an electron in a cyclotron orbit \cite{Peil99}).  To achieve this, one cannot measure the displacement of the resonator \cite{Braginsky80}, but instead must measure the energy directly -- preferably without destroying the quantum state, a so-called quantum non-demolition (QND) measurement.  Whereas the accuracy in continuously measuring two conjugate quantities is limited by the Heisenberg uncertainty principle to the standard quantum limit (SQL) \cite{Braginsky80}, QND measurements allow for continuous measurements of an observable to be taken to arbitrary precision \cite{Santamore04,Hertzberg09, Steinke13, Szorkovszky14}.  Here our interest lies in a QND measurement of the energy, and thereby the number of phonons \cite{Gangat11}.  In an optomechanical system, this is expected to be possible by having strong second order optomechanical coupling \cite{Thompson08,Nunnenkamp10,Huang11,Borkje13}. This has been demonstrated in membrane-in-the-middle Fabry-P\'erot cavities \cite{Sankey10,Lee14}, however it has been pointed out there remains first order coupling between the two optical modes, possibly obscuring QND measurements \cite{Miao09}.

Signal from second order optomechanical coupling, hence measurement of $x^2$, will display mechanical peaks at twice the fundamental frequency.  However, we would also expect that nonlinear transduction of the displacement, $x$, of a mechanical resonator from a nonlinear optical transfer function would also appear at harmonics of the mechanical resonance frequency, as has been observed \cite{Rokhsari05, Huang12,Liu12}.  

\begin{figure}
\includegraphics[width=8.6cm]{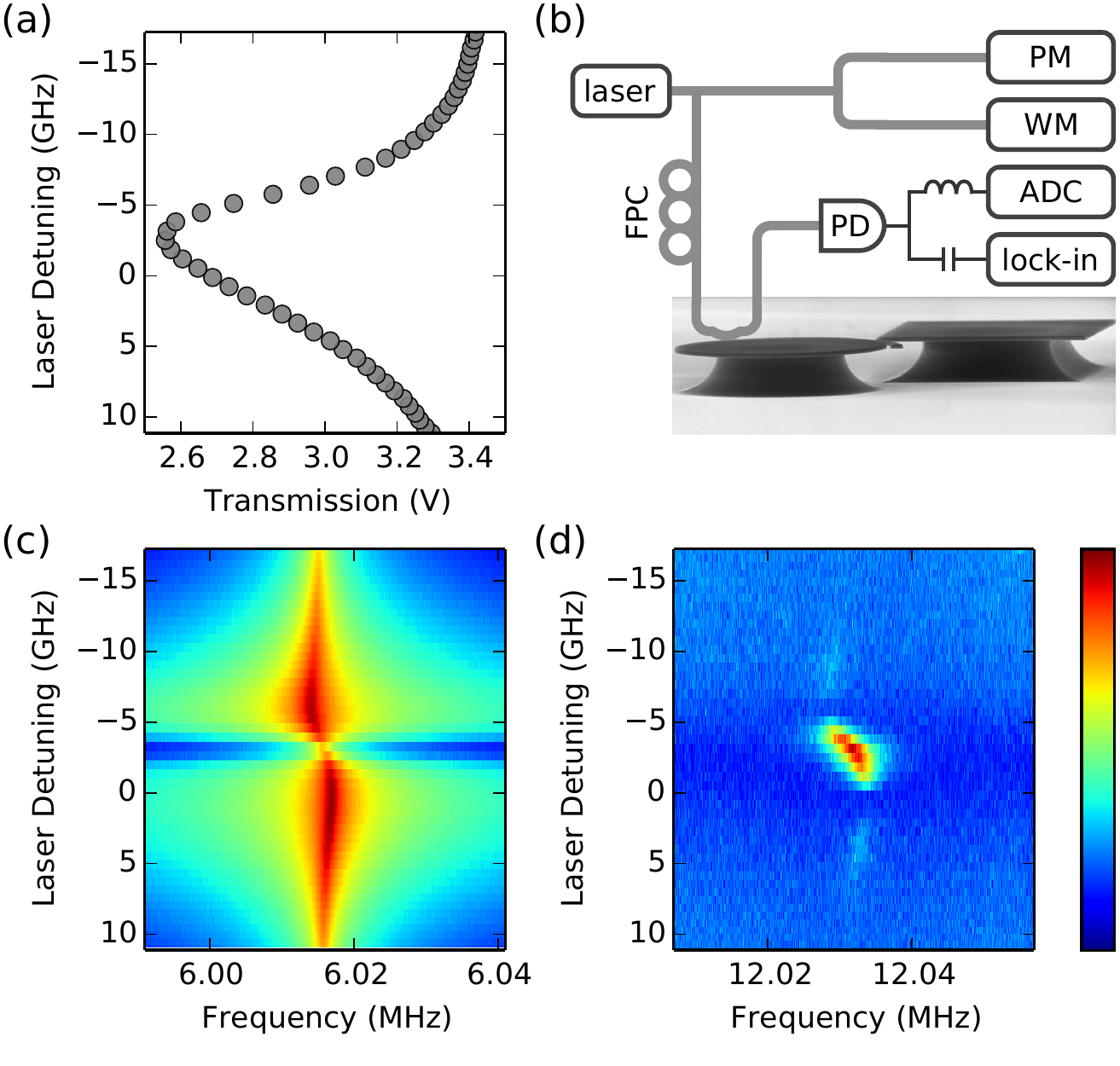}
\caption{\label{fig.1}
(a)  The low frequency transmission ($<1$ kHz) through the optical cavity reveals the optical resonance as a tunable laser is scanned over the optical resonance frequency.  (b)  A schematic of the experiment above a tilted scanning electron microscope (SEM) image of the optomechanical device being measured.  PM - power meter, WM - wavelength meter, ADC - low frequency analog-to-digital converter.  PD - photodetector, FPC - fiber polarization controller.  Scale bar 5 $\mu$m.  Normalized transmission power spectral densities around (c) $1\Omega$ and (d) $2\Omega$.  Frequency doubling indicates nonlinear transduction.  Log-scaled color bar spans from the minimum noise floor ($10^{0}$ in both) up to (c) $10^{5.4}$ and (d) $10^{1}$.
}
\end{figure}

In this Letter we report observation of peaks in the mechanical power spectra at exactly twice the fundamental mechanical frequency, as shown in Fig.\thinspace\ref{fig.1}.  We derive a model for the origin of the harmonic signal, as well as the optical spring effect, from both linear and quadratic optomechanical couplings as a function of laser detuning from the cavity resonance.  We are thus able to determine the coupling contributions to the nonlinear optical transfer function and find second order optomechanical coupling of $\sim$MHz/nm${^2}$, comparable to the membrane-in-the-middle system \cite{Thompson08}.  We also demonstrate the role of second order optomechanical coupling in our signals, providing a framework for enhancing this effect.  Maximizing second order optomechanical coupling while eliminating first order coupling should provide a route to QND measurements at low temperatures.

%%%%%%%%%%%%%%%%%%%
%
%\section{The Optomechanical System}
%
%%%%%%%%%%%%%%%%%%%

The optomechanical cavity being measured is a nanocantilever with effective mass $m = 240$ fg, as described elsewhere \cite{Doolin14}, fabricated on-chip, in the evanescent field of an optical microdisk.  The Hamiltonian for independent optical and mechanical cavities can be written $\hat{H} = \hat{H}_{\rm opt} + \hat{H}_{\rm m}$, where $\hat{H}_{\rm opt} = \hbar \omega_0 \left( \hat{a}^\dagger \hat{a} + 1/2\right)$ and $\hat{H}_{\rm m} = \hbar \Omega_0 \left( \hat{b}^\dagger \hat{b} + 1/2\right)$ are the Hamiltonians of the optical and mechanical resonators.  Here we denote $\omega_0$ and $\Omega_0$ as the optical and mechanical cavity resonance frequencies, and $\hat{a}^\dagger$ ($\hat{b}^\dagger$) and $\hat{a}$ ($\hat{b}$) are the creation and annihilation operators for photons (phonons).  We note that since we will be extending our discussion to the classical regime where the number of quanta in the resonator is large,  we will ignore the ground state contribution to the resonators' energies.

Being within the optical mode volume,  the mechanical resonator's motion is coupled to the optical cavity resonance frequency through shifts in the effective index of refraction.  This can be described to second order as
\begin{equation}\label{eq.coupling}
\omega_0 \rightarrow \omega_0 - G_1 \hat{x} - G_2 \hat{x}^2, \\
\end{equation}
where $G_1 = -\partial \omega_0 / \partial x$ and $G_2 = -(1/2) \partial^2 \omega_0 / \partial x^2$ are the first and second order optomechanical coupling constants.  Therefore
\begin{equation}
\hat{H}_{\rm int} = -\hbar\left(G_1 \hat{x} + G_2 \hat{x}^2 \right)\hat{a}^\dagger \hat{a} 
\end{equation}
is the interaction Hamiltonian to second order.  For a device with symmetric out-of-plane motion in a symmetric  evanescent optical field one anticipates second order optomechanical coupling, with first order coupling arising from asymmetries in the motion or optical field \cite{Doolin14}.

%%%%%%%%%%%%%%%%%%%%
%         FIG 1
%%%%%%%%%%%%%%%%%%%%

%%%%%%%%%%%%%%%%%%%
%
%\section{Optical Transduction}
%
%%%%%%%%%%%%%%%%%%%

We measure the optical transmission through a tapered optical fiber coupled to the optical resonator, Fig.\thinspace\ref{fig.1}b, in the ``tuned-to-slope'' regime.  As such, the classical field in an optical cavity, $a = \left<\hat{a}\right>$, coupled to one input waveguide carrying field $\sqrt{s\,}e^{-i \omega t}$ and one output waveguide carrying away field $z$, when written in a frame rotating at the source frequency $\omega$, can be modeled as
\begin{equation} \label{eq.oeom}
\dot{a} = -\kappa a + i \Delta a + \sqrt{2 \kappa_{\rm e}s\,},
\end{equation}
where $a$ is normalized such that $a^* a = \left<\hat{n}\right> = n$ is the number of photons in the cavity, $\kappa = \kappa_0 + \kappa_{\rm e}$ describes the total loss rate from the optical resonator both to the output waveguide ($\kappa_{\rm e}$) and elsewhere ($\kappa_0$) \cite{Aspelmeyer13}, $\Delta = \omega - \omega_0$ is the detuning of the source laser frequency ($\omega$) from the cavity frequency ($\omega_0$), and $s$ is the incoming power in photons per second.  Note that we define $\kappa$ as the half width at half max of the optical power resonance, such that the cavity rings down as $e^{-2\kappa t}$.

We restrict our analysis to the sideband unresolved regime where $\kappa \gg \Omega_0$ (for the device presented here $\kappa / \Omega_0 \approx 10^3$), hence the optical fields in the cavity reach steady state in a characteristic time $\tau_{\rm opt} = \pi /\kappa$ much faster than the time scale of the mechanical motion ($\tau_{\rm m} = 2\pi / \Omega_0$). A steady state solution to \eqref{eq.oeom} can be found by setting $\dot{a} = 0$:
\begin{equation} \label{eq.a0}
a = \frac{\sqrt{2 \kappa_{\rm e}s\,}}{\kappa - i \Delta},
\end{equation}
giving the number of cavity photons, $n = a^* a$, as
\begin{equation}
n = \frac{2 \kappa_e s}{\kappa^2} \frac{1}{1 + \delta^2}.
\end{equation}
Here we have introduced $\delta = \Delta / \kappa$, the normalized laser detuning from the cavity resonance in units of $\kappa$.

\begin{figure}
\includegraphics[width=3.4in]{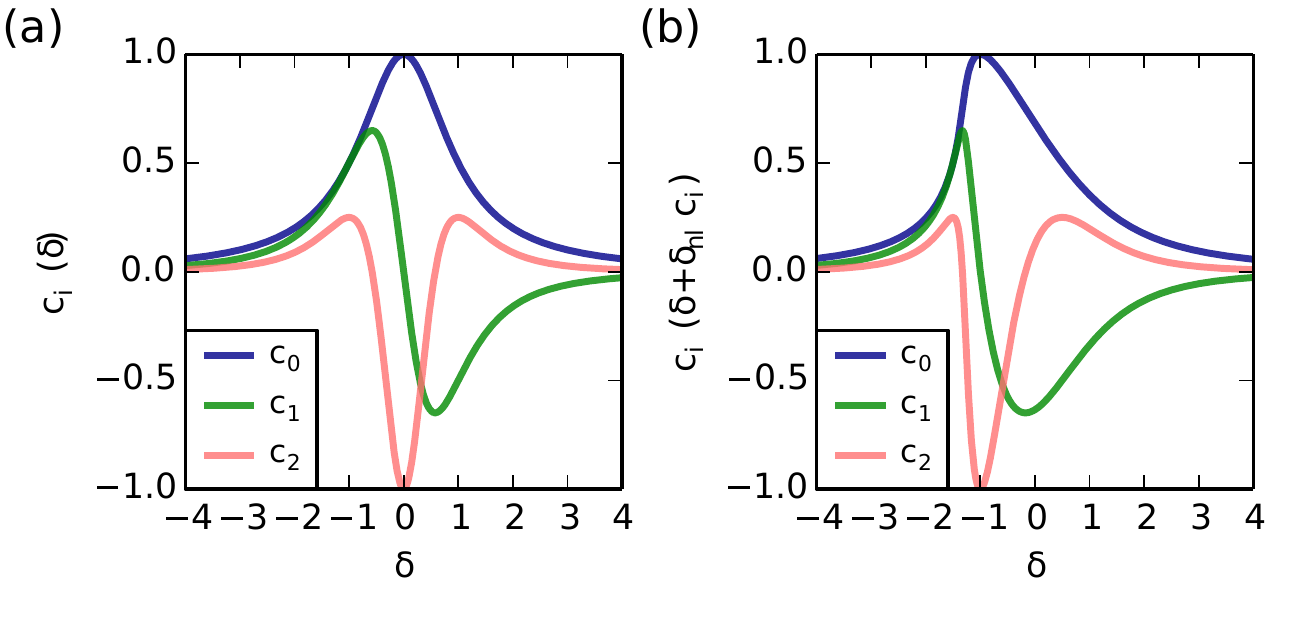}
\caption{\label{fig.2}
(a)  The $c_i$ coefficients are unitless, normalized functions which describe the detuning dependence of the various optomechanical parameters.  By comparing the observed detuning dependence of mechanical spectra with the shape of these $c_i$'s, the origin of the effects can be determined.  (b)  Nonlinear effects due to large optical power in the resonances create asymmetries in the detuning dependence. By adding an additional power dependent detuning these nonlinear effects can be accurately modeled \cite{Barclay05}, as described in the text.
}
\end{figure}

Taylor expanding the Lorentzian detuning dependence of $n$, $c_0(\delta) = (1 + \delta^2)^{-1}$, for small perturbations $u$ around $\delta$, we find
\begin{align}
c_0(\delta + u) &= \frac{1}{1 + (\delta + u)^2} \\
c_0(\delta + u) &= c_0(\delta) + c_1(\delta) u + c_2(\delta) u^2 + O(u^3) \label{eq.c0exp}
\end{align}
where $c_i$ are dimensionless functions of detuning. These are given by
\begin{align}\label{eq.c0}
c_0(\delta) &= \frac{1}{1 + \delta^2} \\
c_1(\delta) &= - \frac{2 \delta}{(1 + \delta^2)^2} \\
c_2(\delta) &= \frac{3 \delta^2 - 1}{(1 + \delta^2)^3} \\ \label{eq.ci}
c_i(\delta) &= \frac{1}{i!} \frac{{\rm d}^i}{{\rm d} \delta ^i} c_0(\delta),
\end{align} 
plotted in Fig.\thinspace\ref{fig.2}a, such that
\begin{equation}\label{eq.optu}
n(\delta + u) \approx n_{\rm max} [ c_0(\delta) + c_1(\delta) u + c_2(\delta) u^2 ],
\end{equation}
where $n_{\rm max} = 2\kappa_e s / \kappa^2$.
Explicitly substituting the coupling of the mechanical motion to the cavity detuning, as given by (\ref{eq.coupling}),
\begin{align}
\delta &\to \delta + \frac{G_1}{\kappa} x + \frac{G_2}{\kappa} x^2, \label{eq.g2pert}
\end{align}
and $G_1 x / \kappa + G_2 x^2 / \kappa$ as $u$ (keeping terms to second order in $x$) we find
\begin{equation}\label{eq.apert}
n \approx \frac{2 \kappa_e s}{\kappa^2}  \left[ c_0 + c_1\frac{G_1}{\kappa} x + \left(c_1\frac{G_2}{\kappa} + c_2\frac{G_1^2}{\kappa^2} \right)  x^2 \right],
\end{equation}
remembering $c_i$ are implicit functions of detuning.

%%%%%%%%%%%%%%%%%%%%
%         FIG 2
%%%%%%%%%%%%%%%%%%%%

Using power conservation, the transmitted field through the optomechanical cavity is
\begin{equation}
z = \sqrt{s\,} - \sqrt{2 \kappa_e} a.
\end{equation}
Substituting in \eqref{eq.a0} for $a$, the power $Z = |z|^2$ detectable at a photodetector is then
\begin{equation}
Z = |z|^2 = s -  \frac{4 \kappa_e \kappa_0 s}{\kappa^2} c_0(\delta).
\end{equation}
Using (\ref{eq.c0exp}),
\begin{equation}\label{eq.Tpert}
Z \approx s  -  \frac{4 \kappa_e \kappa_0 s}{\kappa^2}\left[ c_0 + c_1\frac{G_1}{\kappa} x + \left(c_1\frac{G_2}{\kappa} + c_2\frac{G_1^2}{\kappa^2} \right)  x^2 \right].
\end{equation}

Equation \eqref{eq.Tpert} is the time-series representation of the optical transfer function up to order $x^2$, with three parts: one DC, one oscillating at $1\Omega$, and one oscillating at $2\Omega$.  This can be seen by noting the DC dependence is simply 
\begin{equation}\label{eq.VDC}
Z_{DC} = s  -  \frac{4\kappa_e \kappa_0 s}{\kappa^2}c_0.
\end{equation}
This describes the optical resonance as a function of detuning, as seen in Fig.\thinspace\ref{fig.1}a.
Fourier transforming equation \eqref{eq.Tpert} to linear $x$ and neglecting DC components we find
\begin{equation}\label{eq.V1f}
Z_{\rm 1 \Omega}(\Omega) = -4 s \frac{\kappa_e \kappa_0}{\kappa^2} \frac{G_1}{\kappa} c_1 x(\Omega),
\end{equation}
such that $-4 s \kappa_e \kappa_0 G_1 c_1 / \kappa^3$ is the linear, time-invariant part of the optical transfer function.  Equation \eqref{eq.V1f} describes a typical optomechanical transduction of mechanical signal, as seen in Fig.\thinspace\ref{fig.1}c.

 The remaining nonlinear terms arise from the $x^2$ dependence:
 \begin{equation}\label{eq.V2t}
Z_{\rm 2\Omega}(t) = -\frac{4 \kappa_e \kappa_0 s}{\kappa^2} \left( c_1\frac{G_2}{\kappa} + c_2\frac{G_1^2}{\kappa^2} \right)  x^2(t).
\end{equation}  
Examining $x(t)$ while taking $\Gamma \to 0$ demonstrates the quadratic nature of the spectra;
$x^2(t) \approx x_0 \cos^2 \Omega t = x_0 / 2 \left(1 + \cos 2\Omega t \right)$, 
mixing the $x^2$ signal to $\Omega + \Omega$ and $\Omega - \Omega$ (DC).   Here we neglect the DC signal from the nonlinear transduction, as it will be much smaller than the DC signal from the optical resonance.  These three parts of the optomechanical transduction will be fit to the experimental data to determine the linear and nonlinear optomechanical couplings, $G_1$ and $G_2$.

\begin{figure}
\includegraphics[width=3.4in]{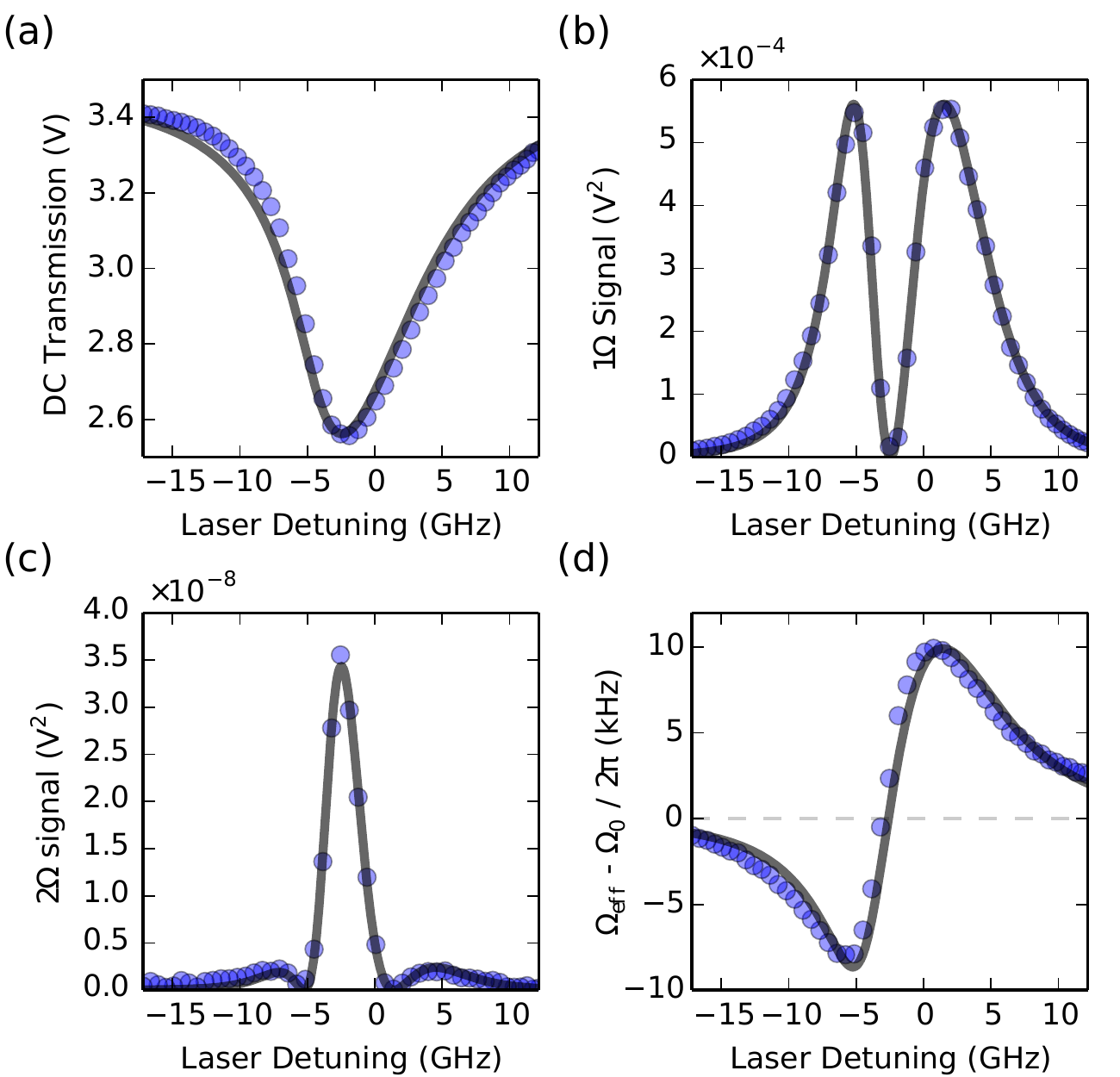}
\caption{\label{fig.3}  Quantitative signals extracted from Fig.\thinspace\ref{fig.1}a, c and d, with respect to laser detuning, $\delta$:
(a) DC optical resonance, (b) $1\Omega$ signal, (c) $2\Omega$ signal, and (d) optical spring effect.  Blue points in (b and c) are found by summing across the mechanical bandwidth.  Grey curves are simultaneous least squares fits to all four signals as explained in the text.
}
\end{figure}

%%%%%%%%%%%%%%%%%%%
%
%\section{Mechanical Back-action}
%
%%%%%%%%%%%%%%%%%%%
While the optical cavity is interacting with the motion of the mechanical resonator, radiation pressure forces provide back action on the resonator's momentum.  These forces can be found classically from the interaction Hamiltonian,
\begin{equation}
F = - \frac{\partial}{\partial x} H_{\rm  int} = \hbar G_1 n + 2 \hbar G_2 n\, x.
\end{equation}
Substituting in our perturbation for $n$ from \eqref{eq.apert} and putting these forces into the equations of motion for a thermally driven damped harmonic oscillator and retaining only force components up to linear in $x$, we find
\begin{eqnarray}
m \ddot{x} + m \Gamma \dot{x} &+& m \Omega_0^2 x =  F_{\rm th} + \hbar n_{\rm max} G_1 c_0  \nonumber\\
&+& \frac{2 \hbar \kappa_e s}{\kappa} \left( \frac{G_1^2}{\kappa^2} c_1 + 2 \frac{G_2}{\kappa} c_0 \right) x,
\end{eqnarray}
where $\Gamma$ is the mechanical damping rate and $F_{\rm th}$ represents uncorrelated thermal forces actuating the resonator.  Collecting terms proportional to $x$, we see the radiation pressure forces shift the effective oscillating frequency of the resonator, $\Omega_{\rm eff}$:
\begin{equation}
m \Omega_{\rm eff}^2 = m \Omega_0^2 - \frac{2 \hbar \kappa_e s}{\kappa} \left( \frac{G_1^2}{\kappa^2} c_1 + 2 \frac{G_2}{\kappa} c_0 \right),
\end{equation}
or
\begin{equation}
\Omega_{\rm eff} - \Omega_{0} \simeq -\frac{\hbar \kappa_e s}{m \Omega_0 \kappa} \left( \frac{G_1^2}{\kappa^2} c_1 + 2 \frac{G_2}{\kappa} c_0 \right).
\end{equation}
Importantly, this optomechanical spring effect has dependence on both $G_1$ and $G_2$ -- similar to the optomechanical transduction -- yet has different dependence on detuning, $\delta$, providing a complementary determination of $G_1$ and $G_2$ (Fig.\thinspace\ref{fig.1}c and Fig.\thinspace\ref{fig.3}d).

%%%%%%%%%%%%%%%%%%%
%
%\section{Nonlinear effects on the optical resonator}
%
%%%%%%%%%%%%%%%%%%%

The displacement transduction and optical spring equations given so far have detuning dependence derived from the $c_i$ functions defined above \eqref{eq.c0}-\eqref{eq.ci}, which have symmetric ($c_0$, $c_2$) or antisymmetric ($c_1$) dependence on laser detuning.
However, the observed detuning dependancies (Fig.\thinspace\ref{fig.1}a, c, d) are stretched towards negative detuning as compared with $c_i$.  This effect can be described by a nonlinearity in the optical resonance whereby the cavity resonance frequency depends on the number of circulating photons \cite{Barclay05}.  This can arise from the optical Kerr effect, or from heating of the microdisk resonator causing changes in the index of refraction.  Following the work of Barclay \textit{et al.\thinspace}\cite{Barclay05}, only one additional parameter is needed to account for this, $\delta_{\rm nl}$, which is a power dependent shift to the detuning.  This can be described mathematically as
\begin{equation}
n = \frac{2 \kappa_e s}{\kappa^2} \frac{1}{1 + \left(\delta + \delta_{\rm nl} n \right)^2},
\end{equation}
which can be numerically solved at each detuning for $n$.  This shift in resonance frequency,  $\delta_{\rm nl} n$, is added to the laser detuning to compensate for the asymmetric shifts in resonance frequency.  In Fig.\thinspace\ref{fig.2}b we show example $c_i(\delta + \delta_{\rm nl} c_i )$ functions.

%%%%%%%%%%%%%%%%%%%%
%         FIG 3
%%%%%%%%%%%%%%%%%%%%

%%%%%%%%%%%%%%%%%%%
%
%\section{The Experiment}
%
%%%%%%%%%%%%%%%%%%%

To collect data, 1590 nm light from a tunable diode laser is transmitted through the optomechanical cavity, coupled via a tapered-dimpled fiber \cite{Michael07,Hauer14} touching the microdisk, and collected on a photodetector.  The voltage output of the photodetector was simultaneously measured with a low frequency analog-to-digital converter and a 50 MHz digital lock-in amplifier performing heterodyne downconversion to allow low sample rate measurements of the signal within a $\sim 60$ kHz bandwidth of both the $1 \Omega$ and $2 \Omega$ signals.  The tunable laser was scanned across the optical resonance with $\approx 3.6$ s of high frequency transmission data recorded for each detuning, while calibrating laser drive frequency with an external wavelength meter.  The power spectral densities (PSDs) \cite{Hauer13} of the 1$\Omega$ and 2$\Omega$ signals were estimated from Fourier transforming the time series data \cite{Bartlett48}.  

The signals at $1\Omega$ and 2$\Omega$ were measured by integrating across the measured PSD bandwidth and subtracting the contribution from the noise floor.  The spectrally white off-resonance noise floor was detuning dependent, and extracted across both the $1\Omega$ and $2\Omega$ signals.  The $1\Omega$ PSD was fit with a damped harmonic oscillator spectrum \cite{Hauer13}, extracting values for $\Omega_{\rm eff}$ (Fig.\thinspace\ref{fig.1}d) and $\Gamma$ (Fig.\thinspace\ref{fig.5}a) at each detuning.  

The power going into the optomechanical cavity, 540 $\mu$W, was calibrated by measuring the laser power before the tapered fiber with a power meter.  The tapered fiber was measured to have near 100\% transmission when not coupling, and scattering losses of 36\% from touching the optical microdisk.  These losses gave excellent agreement to the photodetector's received power and were used to determine $s = 2.8 \times 10^{15}$ photons per second 

%%%%%%%%%%%%%%%%%%%%
%         FIG 4
%%%%%%%%%%%%%%%%%%%%
\begin{figure}
\includegraphics[width=3.4in]{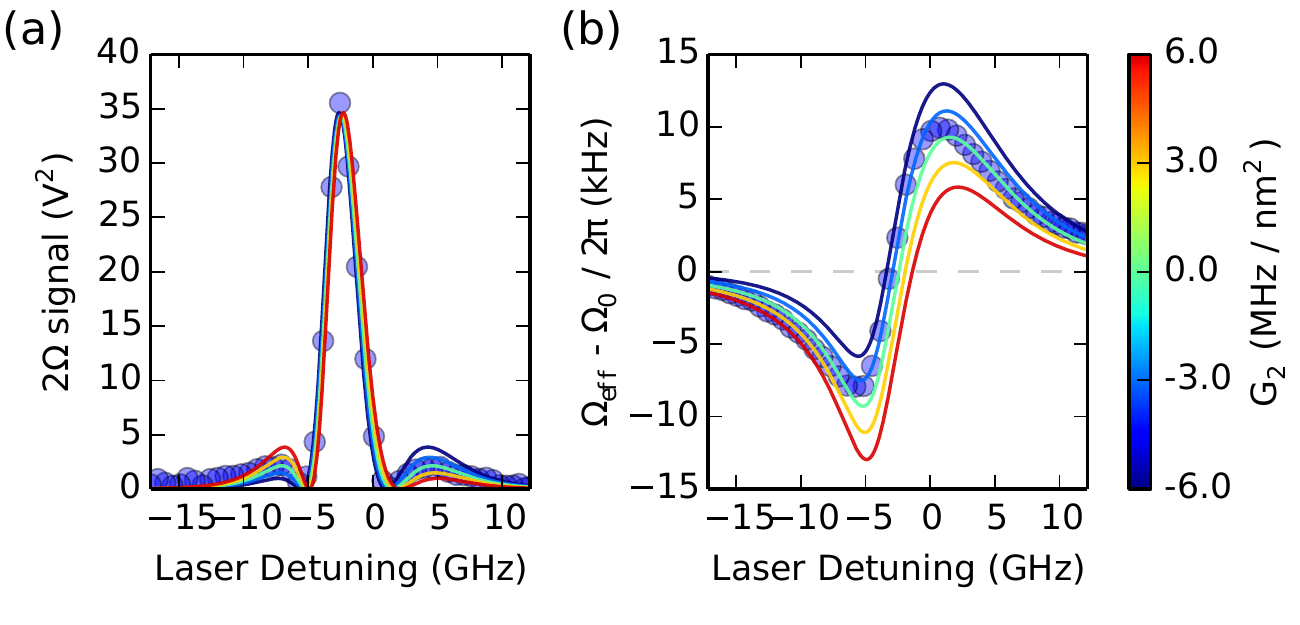}
\caption{\label{fig.4}
Dependence of the (a) $2\Omega$ and (b) optical spring effect signals on the second order optomechanical coupling, $G_2$.  Here $G_2$ transfers spectral weight between detunings.
}
\end{figure}

A nonlinear least squares fit was performed simultaneously to all four datasets presented in Fig.\thinspace\ref{fig.3}, that is, the three parts of the optomechanical transfer function -- DC optical resonance, mechanical signal at $1\Omega$ and at $2\Omega$ -- as well as the optical spring effect.  From the fit the following parameters were extracted: $\kappa = 5.82 \pm 0.02$ GHz, $\kappa_e =  0.42 \pm 0.01$ GHz, $\delta_{\rm nl} \kappa = 5.2 \pm 0.6$ kHz/photon, $G_1 = 458 \pm 2$ MHz/nm, $G_2 = -1.0 \pm 0.6$ MHz/nm$^2$, and $\Omega_0 / 2\pi = 6015.3 \pm 0.3$ kHz.  Errors are standard deviations estimated from the fit covariance.  The least squares algorithm used is only guaranteed to have found a local minimum, however it provides representative numbers and is in reasonable agreement with a previous independent calculation of $G_1$ \cite{Doolin14}.

In the present device the signal at $2\Omega$ is dominated by the contribution to the optomechanical transfer function from the curvature of the optical resonance, that is the term proportional to $c_2 G_1^2 / \kappa^2$.  While this signal is a measurement of $x^2$, it is not appropriate for a QND measurement as the optical resonator continuously introduces back action into the phase, creating uncertainty in $\hat{x}$ at future times.  Only the  contribution from $G_2$ is pertinent to a QND measurement of the energy, and in order to elucidate this we show in Fig.\thinspace\ref{fig.4} the $2\Omega$ signal and the optical spring effect data with varying $G_2$ while keeping all other parameters fixed.  It is interesting to note that while the sign of $G_1$ is irrelevant, the sign of $G_2$ is important.  Specifically, moving from negative to positive values of $G_2$ shifts spectral weight as a function of detuning.
 
   We note that our optomechanical coupling constants $G_1$ and $G_2$ correspond respectively to a single photon to single phonon coupling rate ($G_1 x_{\rm zpf}$) of 35 kHz, and a single photon to two phonon coupling rate $|G_2 x^2_{\rm zpf}|$ of 6 mHz, where $x_{\rm zpf} = \sqrt{\hbar/2m \Omega_0}=76.2$ fm.  We expect that the single photon - two phonon coupling rate should be larger than $\Gamma$ ($\sim1.44$ kHz) to make a continuous measurement of the quantized energy states before decoherence -- not satisfied with the present device, although measurements of phonon shot noise may be possible with weaker coupling \cite{Clerk10}.

%%%%%%%%%%%%%%%%%%%%
%         FIG 5
%%%%%%%%%%%%%%%%%%%%
\begin{figure}
\includegraphics[width=3.4in]{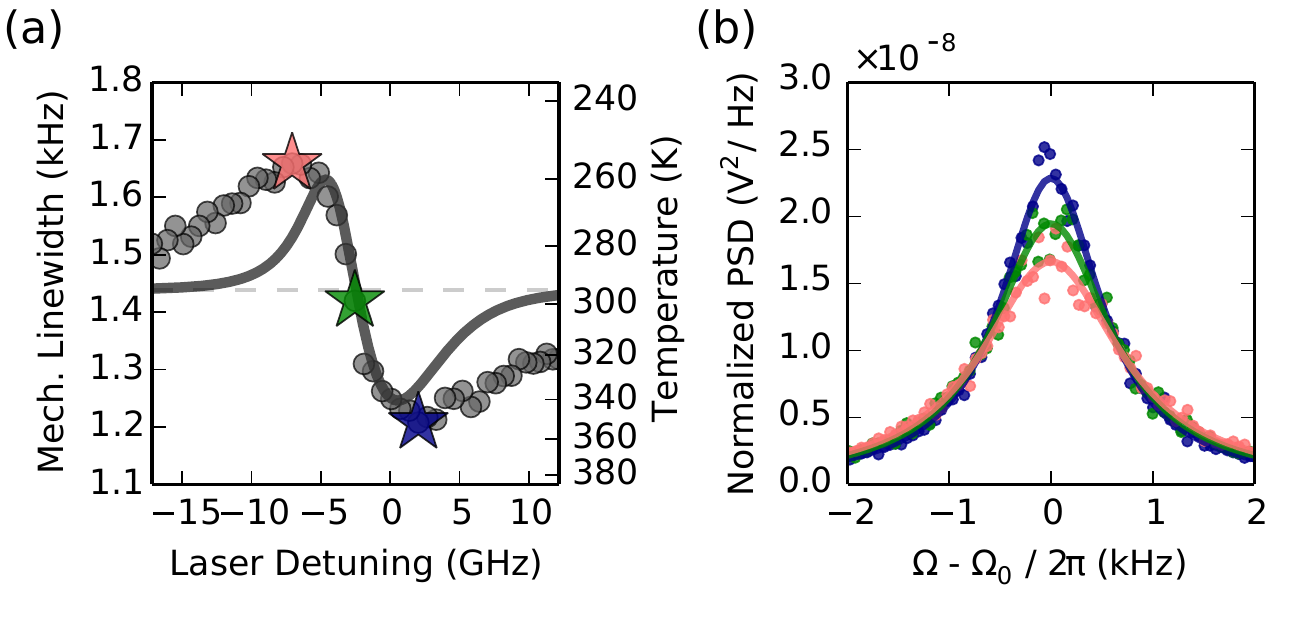}
\caption{\label{fig.5}
Dynamical effects on the mechanical resonator from the optical cavity.  (a) The linewidth of the mechanical resonator is damped and then amplified as the optical detuning is varied from negative (red detuning) to positive (blue detuning).  Curve is the theoretical mechanical linewdith \cite{Aspelmeyer13} using the parameters derived from the nonlinear least squares fit.  (b)  The corresponding power spectral densities at three detunings.  The colors of the data correspond to the stars in panel (a), and are 349 K (red), 297 K (green) and 254 K (blue) \cite{Marquardt08}.
}
\end{figure}

%\section{Dynamical Backaction}

Finally, our stationary regime model does not account for dynamical backaction to the mechanical spectrum, as the optical resonance is always in its steady state.  Nonetheless, backaction from light-induced forces in the sideband unresolved regime is expected \cite{Marquardt08, Aspelmeyer13}.  Analysis of the mechanical linewidth as a function of detuning from the $1\Omega$ signal (from Fig.\thinspace\ref{fig.1}c) reveals non-negligible optomechanical damping and amplification, presented in Fig.\thinspace\ref{fig.5}.  The theoretical curve for optomechanical damping, using the fit parameters determined in Fig.\thinspace\ref{fig.3}, is given in Fig.\thinspace\ref{fig.5} with reasonable agreement \cite{Marquardt08,Aspelmeyer13}.  As a result the mechanical mode is heated (cooled) from 297 K to 349 K (254 K). 

%\section{Conclusion}

Motivated by the search for experimentally realizable approaches to QND measurements of a nanomechanical resonator's energy, we have developed a method to separate nonlinear transduction of first order optomechanical coupling from second order optomechanical coupling.  Both transduction mechanisms give rise to frequency doubling in the mechanical spectrum,  however the detuning dependence in both the $2\Omega$ power spectrum and the optomechanical spring effect allow determination of the optomechanical coupling constants.  
Fitting our experimental data to these models reveals a second order coupling, $G_2$, of MHz/nm$^2$.  Enhancing this second order coupling, and eliminating the first order optomechanical coupling, through fabricating a fully symmetric device in both mechanical motion and evanescent optical field, should provide an approach to QND measurements of phonon number, as well as exotic phenomena such as quantum superpositions of nanomechanical resonators \cite{Tan13}.

% figures should be put into the text as floats.
% Use the graphics or graphicx packages (distributed with LaTeX2e)
% and the \includegraphics macro defined in those packages.
% See the LaTeX Graphics Companion by Michel Goosens, Sebastian Rahtz,
% entire page. There is no need to do explicit centering.

%\begin{acknowledgments}
We would like to thank our funding sources: the University of Alberta; the Canada Foundation for Innovation; the Natural Sciences and Engineering Research Council of Canada; Alberta Innovates Technology Futures; and the Alfred P. Sloan Research Foundation.  We also thank K.S.D. Beach, J. Teufel, D. Chang and F. Marquardt for useful suggestions.
%\end{acknowledgments}

% Create the reference section from 2f.bib
\bibliography{2f}

\end{document}